\documentclass[usenatbib]{mnras}

\usepackage{newtxtext,newtxmath}
\usepackage{ae,aecompl}
\usepackage[usenames]{xcolor}

\usepackage[T1]{fontenc}    % use 8-bit T1 fonts
\usepackage{booktabs}       % professional-quality tables
\usepackage{amsfonts}       % blackboard math symbols
\usepackage{nicefrac}       % compact symbols for 1/2, etc.
\usepackage{graphicx}
\usepackage{amsmath}
\usepackage{comment}
\usepackage{mathtools}
\usepackage{tabularx}
\usepackage{xfrac}
\graphicspath{ {./images/} }

\DeclarePairedDelimiter\abs{\lvert}{\rvert}%
\DeclarePairedDelimiter\norm{\lVert}{\rVert}%   

\makeatletter
\let\oldabs\abs
\def\abs{\@ifstar{\oldabs}{\oldabs*}}
\let\oldnorm\norm
\def\norm{\@ifstar{\oldnorm}{\oldnorm*}}
\makeatother

\newcommand{\kmax}{k_{\mathrm{max}}}
\newcommand{\Pk}{\mathcal{P}_k}
\newcommand{\Pnorm}{P_{\mathrm{norm}}}
\newcommand{\Pkpred}{\mathcal{P}_k^{\mathrm{pred}}}
\newcommand{\Pktrue}{\mathcal{P}_k^{\mathrm{true}}}
\newcommand{\hMpc}{\,h\,\mathrm{Mpc}^{-1}}
\newcommand{\Nlos}{N_{\mathrm{los}}}
\newcommand{\Nin}{N_{\mathrm{in}}}
\newcommand{\Nout}{N_{\mathrm{out}}}
\newcommand{\Npoints}{N_{\mathrm{points}}}

\title[Reconstruction of the Power Spectrum using ML]{Reconstruction of the Density Power Spectrum from Quasar Spectra using Machine Learning}

\author[Han Veiga et al.]{
Maria Han Veiga$^{1}$\thanks{Contact e-mail: \href{mailto:mhanveig@umich.edu}{mhanveig@umich.edu}}\href{https://orcid.org/0000-0002-7562-7014}{\includegraphics[scale=0.5]{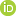}},
Xi Meng$^{2}$\href{https://orcid.org/0000-0002-8276-4164}{\includegraphics[scale=0.5]{figures/orcid.png}},
Oleg Y. Gnedin$^{2}$\href{https://orcid.org/0000-0001-9852-9954}{\includegraphics[scale=0.5]{figures/orcid.png}},
Nickolay Y. Gnedin$^{3}$,
Xun Huan$^{4}$\href{https://orcid.org/0000-0001-6544-2764}{\includegraphics[scale=0.5]{figures/orcid.png}}
\\
$^1$ Michigan Institute for Data Science, University of Michigan, Ann Arbor, MI 48109, USA \\ 
$^2$ Department of Astronomy, University of Michigan, Ann Arbor, MI 48109, USA\\
$^3$ Fermi National Accelerator Laboratory, Batavia, IL 60510, USA\\
$\phantom{^3}$ Kavli Institute for Cosmological Physics, The University of Chicago, Chicago, IL 60637 USA\\
$\phantom{^3}$ Department of Astronomy \& Astrophysics, The University of Chicago, Chicago, IL 60637 USA
\\ 
$^4$ Department of Mechanical Engineering, University of Michigan, Ann Arbor, MI 48109, USA
}

\date{Submitted \today}
\pubyear{2021}

\begin{document}
\label{firstpage}
\pagerange{\pageref{firstpage}--\pageref{lastpage}}
\maketitle

\begin{abstract}
We describe a novel end-to-end approach using Machine Learning to reconstruct the power spectrum of cosmological density perturbations at high redshift from observed quasar spectra. State-of-the-art cosmological simulations of structure formation are used to generate a large synthetic dataset of line-of-sight absorption spectra paired with 1-dimensional fluid quantities along the same line-of-sight, such as the total density of matter and the density of neutral atomic hydrogen. With this dataset, we build a series of data-driven models to predict the power spectrum of total matter density. We are able to produce models which yield reconstruction to accuracy of about 1\% for wavelengths $k \leq 2\hMpc$, while the error increases at larger $k$. We show the size of data sample required to reach a particular error rate, giving a sense of how much data is necessary to reach a desired accuracy. This work provides a foundation for developing methods to analyse very large upcoming datasets with the next-generation observational facilities.  
\end{abstract}


\section{Introduction}

New major observational facilities in the coming decade -- in particular the James Webb Space Telescope and the next generation 30-meter class telescopes such as the GMT, TMT, and E-ELT -- will produce an explosion in the quantity and quality of observational data for thousands of quasars. Current samples of $\sim100$ quasars already known at redshift $z>6$ \citep{wang2019} will increase to several thousand. The largest progress will occur in quasar absorption spectroscopy, which allows a unique and particularly powerful probe of quasar environment. Development of advanced numerical methods is required to take full advantage of these upcoming datasets.

Recently, machine learning (ML) methods have begun to be introduced in astronomical applications. For example, unsupervised deep learning methods have been used to identify and discover exoplanets \citep{Shallue_2018,McCauliff_2015}, or classification models to generate accurate galaxy morphology catalogues \citep{sanchez2018}, predict structure formation \citep{He13825}. For an non-exhaustive review of ML methods in astronomy please refer to \citep{brunner2010,carleo2019}.

In this paper we show how ML methods can be used to relate the clustering of transmission spikes in quasar absorption spectra to the clustering of matter along the observer's line-of-sight (LoS). To make this possible, we use data from the current state-of-the-art numerical simulations of galaxy formation \citep{gnedin_14}. The outputs of these simulations allow us to produce a large sample of synthetic quasar spectra, for which we know the corresponding true properties of cosmic gas, responsible for each spectral feature in the simulated spectra. The complexity in inferring gas properties from spectral data lies in the multitude of physical processes taking place around quasars and the non-trivial relation between the locations in physical space and specific wavelengths of spectral features (the so-called ``redshift space''). We will leverage supervised learning to discover these identifying features and predict matter properties from spectral data.

The absorption spectra of quasars outside of their host galaxies provide clues on the interplay of cosmic gas flows, formation of stars inside nearby galaxies, and the clustering of matter on cosmological scales. Previous efforts to explore the relation between the quasar absorption spectra and clustering of matter along the quasar line of sight focused primarily on cross-correlating absorption spectra with galaxies as tracers of the full matter distribution \citep{kakiichi2018,becker2018,meyer2019,garaldi2019}.

Our primary goal is to go beyond a simple cross-correlation analysis and to construct mappings from the synthetic quasar spectra (``input'') to properties of matter along the observer's LoS (let us call this ``labels''). The space of potential labels is large: thermodynamic and ionization properties of cosmic gas at each location in space, its chemical composition, distances to nearest galaxies of various masses and star formation rate, duty cycle of quasar activity, dynamical state of the host galaxy, etc. The connections between any of these labels and the input are currently unknown. We seek to build these mappings using deep learning, in order to leverage the very large dataset of quasar spectra and matter properties that can be produced from the galaxy formation simulation.

In this paper we report the initial effort in this direction. We analyze one simulation output at redshift $z\approx 6$ and post-process it to produce a set of 100,000 synthetic absorption spectra. Each LoS samples a region of about 200 comoving $h^{-1}$ Mpc in length. As a first step, we aim to reconstruct the matter density power spectrum (PS) along an observed LoS in the quasi-linear regime on scales 1-100 comoving Mpc. 

The rest of the paper is organized as follows. We describe the data generation procedure in \autoref{sec:synt_spectra}, followed by the problem formulation and ML set-up in \autoref{sec:method}. The results are presented in \autoref{sec:results} with discussion in \autoref{sec:discussion}, and finally conclusions in \autoref{sec:conclusion}.

\section{Synthetic spectra}
\label{sec:synt_spectra}

The first step in our exploration is to create a large dataset of synthetic quasar spectra with known underlying line-of-sight matter properties. To produce such a dataset, we use state-of-the-art numerical simulations of galaxy formation performed with the Adaptive Refinement Tree (ART) code \citep{kravtsov1999,kravtsov2002,rudd2008}. These simulations rely on the Adaptive Mesh Refinement algorithm to boost spatial and temporal resolution for both gas dynamics and gravity calculations. The ART code includes many physical processes that are critically important for modeling the formation of galaxies and quasars. The basic components are gravitational dynamics of dark matter and stars, and fluid dynamics of cosmic gas. The ART code computes the thermodynamics of cosmic gas in a manner that is fully self-consistent, without the commonly made assumption of thermodynamic equilibrium. It tracks the abundances of heavy elements released by stellar winds and supernovae and models the non-equilibrium chemistry network of molecular hydrogen, which is needed to identify star-forming regions within galaxies. The ART code includes some of the most advanced modeling of stellar feedback: momentum ejection due to radiation pressure and stellar winds of young stars, ionizing radiation of massive stars, thermal energy of supernova explosions, and their kinetic feedback deposited directly into gas cells and into driving supersonic turbulence.

The ART code accounts for the radiative transfer of ionizing and ultraviolet radiation, which is critically important for proper modeling of the interstellar and intergalactic gas. The local radiation flux affects ionization states of the chemical elements, heating of cold gas, formation and dissociation of molecular clouds, and the radiation pressure on gas and dust. Modeling of radiative transfer will allow us to follow the propagation of ionizing radiation from the quasar -- a critical advantage of our simulations in modeling the structure of the proximity zones. The primary data source for the proposed work are the numerical simulations with the ART code from the {\em Cosmic Reionization on Computers} project \citep{gnedin_14}.

In the present analysis we use the output of one simulation at redshift $z\approx 6$.
The size of the comoving simulation box is $80\, h^{-1}$~Mpc, which is not long enough to match the range of observed spectra in the quasar vicinity. Taking advantage of the periodic boundary conditions in the simulation box, we wrapped a given line 2.5 times around the box, to obtain full line length $L = 200\, h^{-1}$~Mpc. Such wrapping has been shown to result in negligible artifacts induced by periodic boundary conditions \citep{dallaglio2010}. We sample each line-of-sight with 10240 resolution elements and the spatial comoving resolution $\Delta x \approx 0.0195\, h^{-1}$~Mpc. In total, we generated 100,000 synthetic line-of-sight spectra.

In addition, for each LoS spectrum, we store the information on the total density of matter $\rho(x)$, the density of neutral atomic hydrogen $\rho_{\rm HI}(x)$, the gas temperature $T(x)$, and the gas velocity $u(x)$. Using these quantities we can compute the optical depth at a wavelength $\lambda$ as an integral along the line:
\begin{equation}
  \tau(\lambda) = \int \rho_{\rm HI} \frac{\sigma_0 \, c}{\sqrt{\pi}\, m_H\, b(T)} \exp{\left[-\frac{(u_{\lambda}-u)^2}{b(T)^2}\right]} \frac{dx}{1+z},
  \label{eq:optical_depth}
\end{equation}
where $\sigma_0 = 4.5\times 10^{-18}$~cm$^2$ is the cross-section for resonant Ly$\alpha$ absorption, $b(T) = (2kT/m_H)^{1/2}$ is the Doppler parameter, $m_H$ is the mass of hydrogen atom, and $c$ is the speed of light \citep{gnedin_16}. The dependency on wavelength $\lambda$ enters through $u_\lambda$:
\[ \frac{u_\lambda(x)}{c} = \frac{\lambda}{\lambda_{\rm em}} - 1 - z(x) \]
where $\lambda_{\rm em}$ is the rest-frame emission wavelength. In this work we focus on the normalized absorption flux, $F = e^{-\tau}$, without explicitly modeling the intrinsic quasar emission spectrum.

Given a data field along the line-of-sight (such as transmitted flux or the density) we generate the 1-dimensional perturbation power spectrum along that LoS. The 1D power spectrum $P_k^{1D}$ is related the commonly used 3D version $P^{3D}_k$ via an integral over all higher wavenumbers \citep{lumsden_etal89}:
\begin{equation}
  P_k^{1D} = \int_k^\infty P^{3D}_{k^\prime} \frac{k^\prime}{2\pi} dk^\prime.
  \label{eq:pk1d}
\end{equation}
It incorporates information about all small scales. The 1D PS of the transmitted quasar flux was first analyzed using cosmological simulations by \citet{croft_etal98}.

\section{Methodology}
\label{sec:method}

\subsection{Problem overview}
\label{sec:problem_of_interest}

We are interested in the following inverse problem: given an observed flux spectrum $F(\lambda)$, estimate the fluid quantities that produced this flux. As a first step, we seek to build a mapping from $F$ to the power spectrum of matter density $\rho$. In the context of this mapping, flux is the \emph{input}, and density is the \emph{output}. We approach this goal by training a deep neural network (DNN) model in a supervised learning manner using the simulation datasets. 

\subsection{Data transformation}
\label{ss:data_transform}

Compact high-density regions (halos, filaments) may contribute disproportionately to the 1D power spectrum. In Appendix~\ref{sec:appendix} we show examples of three lines of sight that result in very different shape and normalization of $P_k^{1D}$ depending on whether they happen to pass close to a high enough density region or not.

To reduce this variation among individual lines, and to attribute larger importance toward regions of low density (and therefore higher transmission flux), we study three transformations of the matter density along a given line-of-sight. Given the overdensity $\delta(x) \equiv \rho(x)/\bar{\rho}-1$, where $\bar{\rho}$ is the global average matter density of the universe at the studied epoch, we consider the following transformations:
\begin{enumerate}
    \item Linear transformation: $t(\rho) = \delta$
    \item Logarithmic transformation: $t(\rho) = \ln{(1+\delta)}$ 
    \item Inverse transformation: $t(\rho) = 1 - (1+\delta)^{-1}$.
\end{enumerate}
All three transformations converge to $t(\rho) \approx \delta$ when $\delta\ll 1$. Nevertheless, as shown in Appendix~\ref{sec:appendix}, the logarithmic and inverse transformations significantly reduce the variation among individual lines-of-sight.

The performance of the three choices within the DNN model will be presented in \autoref{sec:results}. The transformed density field now takes the role of the {\it output}.

\subsection{Data representation}
\label{sec:datarep}

We then represent the signals of $F$ and $t(\rho)$ via truncated Fourier expansion:
\begin{equation}
    f(x) \approx \sum_{\alpha=0}^{N(k_{\rm max})} \hat{f}_{\alpha} \, \exp(-i 2\pi \alpha x),
\end{equation}
where $\hat{f}_\alpha \in\mathbb{R}$ denotes linear Fourier coefficients and $N(k_{\rm max})$ the number of terms considered. We select $N(k_{\rm max})$ based on the maximum wavenumber $k_{\rm max}$ we are interested in studying, through the following relation:
\begin{equation}
\label{eq:nk_k}
    N(k_{\rm max}) = \frac{L}{2\pi} k_{\rm max},
\end{equation}
where $L=200\,h^{-1}$~Mpc is the largest length-scale from the simulation domain.

\begin{figure}
  \includegraphics[width=\linewidth]{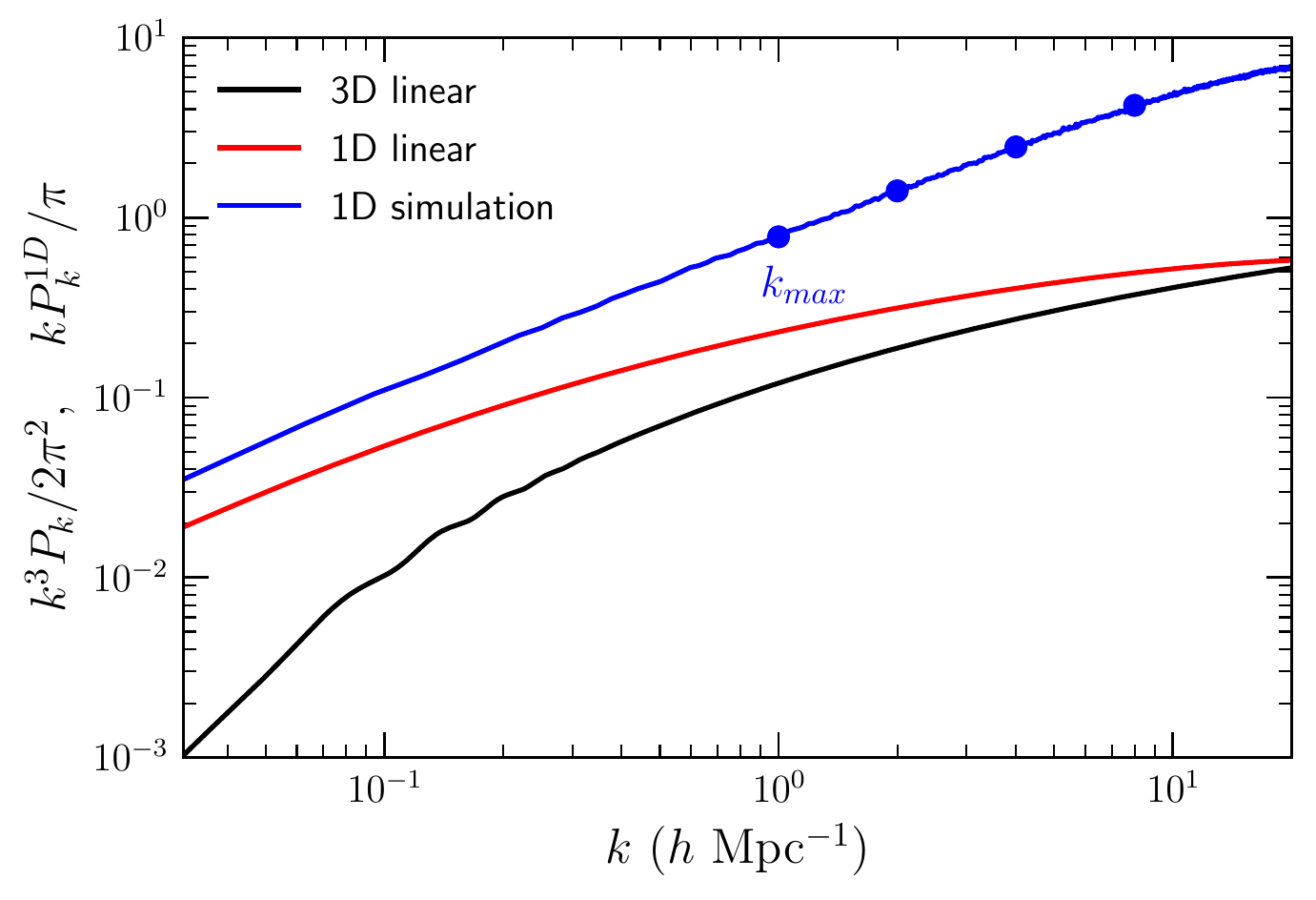}
  \vspace{-5mm}
  \caption{An illustration of the linear power spectrum of cosmological perturbations in one and three dimensions at the epoch $z\approx 6$ studied here. In both cases we show the dimensionless mass perturbation: $k^3 P_k/2\pi^2$ (black line) and $k P_k^{1D}/\pi$ (red line). For comparison we also show the average 1D power spectrum along all lines of sight in the analyzed simulation output (blue line). Circles mark the choices of the limiting wavenumber $\kmax$, which corresponds to the transition to non-linear regime.}
  \label{fig:fourierapprox}
\end{figure}

\autoref{fig:fourierapprox} shows that at wavenumbers $k \gtrsim 1 \hMpc$, the cosmic perturbations are already in a non-linear regime. The residual mean square (rms) mass fluctuation on scale $k$,
$$ \Delta_1 \equiv \frac{k P_{k}^{1D}}{\pi}, $$
reaches $\Delta_1 (k,z) \approx 0.78$ for $k = 1\hMpc$, and $\Delta_1 (k,z) \approx 4.2$ for $k = 8\hMpc$.
Thus this range of wavenumbers includes a mildly non-linear regime. As $\kmax$ is increased, we expect it will become harder to predict the PS from quasar spectra because of potential contribution of local galactic sources to ionization of hydrogen.

Our simulation data has two components: the density field and the flux. The density field has a length of $L$, sampled uniformly by $N=10240$ points. The flux fields do not have the same length, so in order to keep their spectral representation consistent, we truncate all fluxes to the smallest common length. With this procedure, the flux is represented by $N_F = 2800$ points. 

After performing Fourier transform on the flux and density signals, we can extract their respective power spectra for each data pair $j$:
\begin{align}
\label{eq:input_transform}
   F_j &= \left( |\hat{F}_0|^2 , \dots, |\hat{F}_{N(k_{\rm in})}|^2 \right)^{T} \\
\label{eq:output_transform}
   t_j &= \left( |\hat{t}_1|^2 , \dots, |\hat{t}_{N(\kmax)}|^2 \right)^{T}.
\end{align}
Here $k_{\rm in}$ and $\kmax$ denote the maximum wavenumber used to express the input and output, respectively. In our analysis we adopt $k_{\rm in} = 8 \kmax$\footnote{Except for $\kmax=8\hMpc$, where we take $k_{\rm in} = 4 \kmax$ due to the limited number of the input modes.}, chosen to ensure that there is enough information in the input to inform the output. Other choices for this ratio can also be adopted, or one can also keep the input size $N(k_{\rm in})$ fixed throughout the experiments.

Note that we omit the $0$-th mode on purpose in \autoref{eq:output_transform}. The $0$-th mode corresponds to the mean value of the density, and it cannot be recovered from the observed flux as it is completely degenerate with the value of the ionizing background. We study the power of the Fourier modes instead of the Fourier coefficients themselves because the former leads to a translation-invariant representation. 

Analyzing the variable relations line-by-line would be noisy and difficult. We thus consider taking small groupings of the lines in an effort to average out this variability. In this setup, each group contains $N_g$ independent LoS and we compute the average of their power spectra:
\begin{align} 
\label{eq:in_out_groupped_transform}
   F_g(k) = \frac{1}{N_g} \sum_{j=1}^{N_g} F_j(k), \qquad
   t_g(k) = \frac{1}{N_g} \sum_{j=1}^{N_g} t_j(k).
\end{align}
We will also vary the grouping size to study the extent of prediction performance improvement as a result of the reduced statistical variation.

\begin{figure*}
  \includegraphics[width=\textwidth]{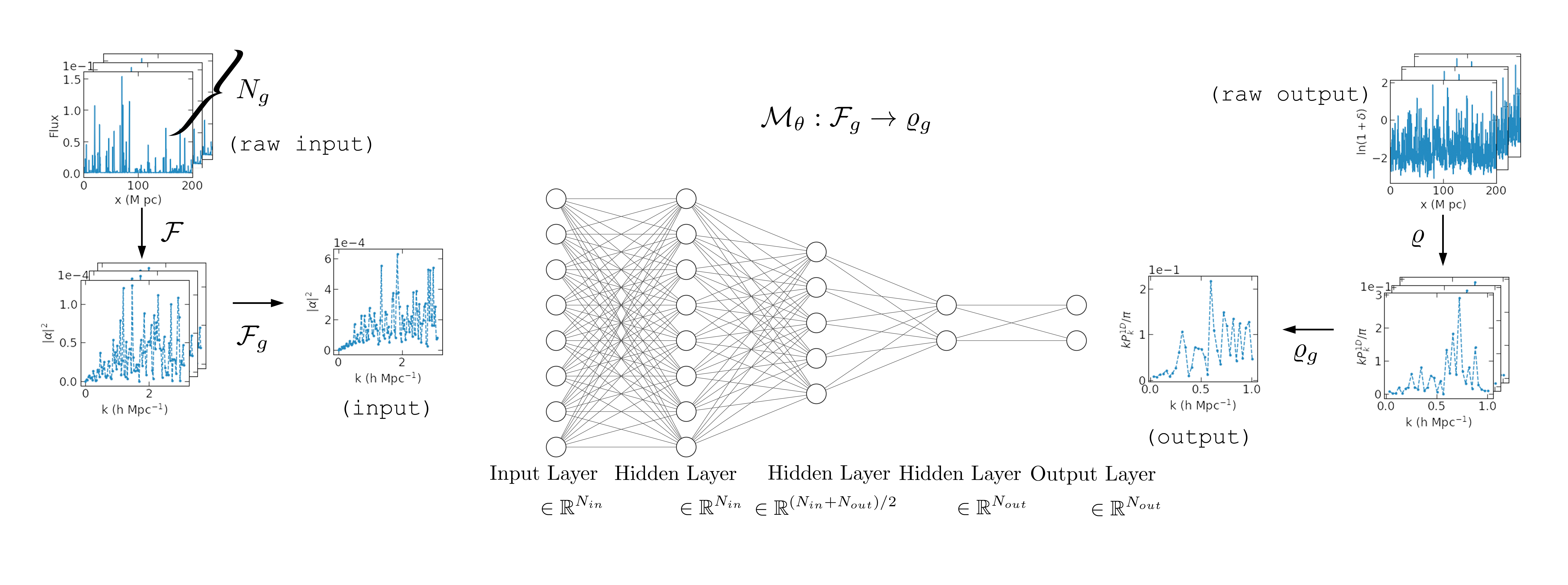}
  \vspace{-9mm}
  \caption{Schematic of the end-to-end inverse problem pipeline. On the left-hand side, the raw signal and its representation (feature) are shown, which will be the input for the neural network. In the middle, neural network structure is shown, depending on the sizes of the input and output vectors. In this diagram, we are setting $\Nin = 8$ and $\Nout = 2$, whereas in reality $\Nin$ is typically 8 times $\Nout$, and $\Nout$ depends on the maximum wavelength $\kmax$ that we are interested in studying (for example, $\kmax = 1\hMpc$ leads to $\Nout = 32$). The right-hand side shows the output of the neural network, as well as the output generation from the raw data, which we use for model training.}
  \label{fig:nn_graphic}
\end{figure*}

\subsection{Deep Neural Network model}
\label{sec:network}

Our inverse problem entails learning a mapping
\begin{equation}
 \label{eq:model2}
 \mathcal{M}: \mathcal{F} \to \tau
\end{equation}
where $\mathcal{F} \subset \mathbb{R}^{\Nin}$ is the vector space containing elements $F_g$, and $\tau \subset \mathbb{R}^{\Nout}$ is the vector space containing elements $t_g$. Here $\Nin = N(k_{\rm in})+1$, and $\Nout = N(\kmax)$.

We approximate $\mathcal{M}$ with $\mathcal{M}_{\theta}$ using a densely-connected DNN (i.e., multilayer perceptron) of the form:
\begin{equation}
\label{eq:ann_model}
    \mathcal{M}_{\theta}(F_g) := h_N(...(h_0(A_0 F_g + b_0))),
\end{equation} 
where $h_i$ are nonlinear activation functions, and $\theta=\{A_i,b_i\}$ collects all trainable parameters (weights and bias parameters) of the DNN. Note that the size of the output (dimension of $\tau$) will depend on the maximal wavenumber $\kmax$ we are interested in predicting. Because of this we consider the following 5-layer architecture, shown schematically in \autoref{fig:nn_graphic}:
\begin{itemize}
    \item An input layer of size $\Nin$ accepting vector $F_g$.
    \item A first hidden layer also with $\Nin$ neurons.
    \item A second hidden layer with ${(\Nin+\Nout)/2}$ neurons.
    \item A third hidden layer with $\Nout$ neurons.
    \item An output layer producing a prediction vector of size $\Nout$.    
\end{itemize}
This architecture was chosen after empirically observing that, when using too few degrees of freedom, the model would under-predict the output variation, and the performance of the network decreased with shallower networks. Rectified linear units (ReLU)\footnote{This activation function is chosen because it is simple and more computationally efficient when compared to hyperbolic tangent or sigmoid. Although there are other suitable activation functions, typically modifications to ReLU (e.g., leaky ReLU, parametric ReLU or randomised leaky ReLU), the empirical improvement on the performance of these is not significant \citep{xu2015}.} are employed for all activation functions $h_i$ of hidden layers, and linear activation is used for the output layer. 

In addition, we normalise the inputs and outputs. For the inputs, we apply a standard scaling relation on each feature (component-wise):
\begin{equation}
\label{eq:standard_escaling}
    F_{\mathrm{in},\alpha} = \frac{F_\alpha - \mu_\alpha}{\sigma_\alpha}, \quad \alpha = 1, \cdots, \Nin,
\end{equation}
where $\mu_\alpha$ is the sample average and $\sigma_\alpha$ is the sample standard deviation of the $\alpha$-th feature across the data points. We are careful to compute these estimates only from the training data to avoid data leakage.

For the outputs, we simply divide the data by a fixed normalization constant, $\Pnorm$:
\begin{equation}
   t_{\mathrm{out}} = \frac{t_{g}}{\Pnorm}.
\end{equation}
This constant is derived from the data and corresponds to the average magnitude of the power spectrum, up to wavenumber $4 \hMpc$. We define it up to a rescaling factor of order unity as
\begin{equation}
  \label{eq:normalisation_output}
   \Pnorm = c \bar{P}, \quad
   \bar{P} = \frac{1}{\Npoints \, N(4)} 
   \sum_{k}^{N(4)} \sum_{l=1}^{\Npoints} t_{g,l}(k).
\end{equation}
The value of $\bar{P}$ remains fixed for the entirety of all numerical experiments. We will investigate the rescaling factor, expected to be $c\sim 1$, as a hyperparameter. This normalization reduces the output range to be of order unity and is similar to standard scaling. We have tested that such a normalisation is extremely important for the performance of the network, as it keeps the input and output values within a small range. Without the appropriate normalisation, the network often fails.

\begin{figure}
\centering
  \includegraphics[width=0.35\textwidth]{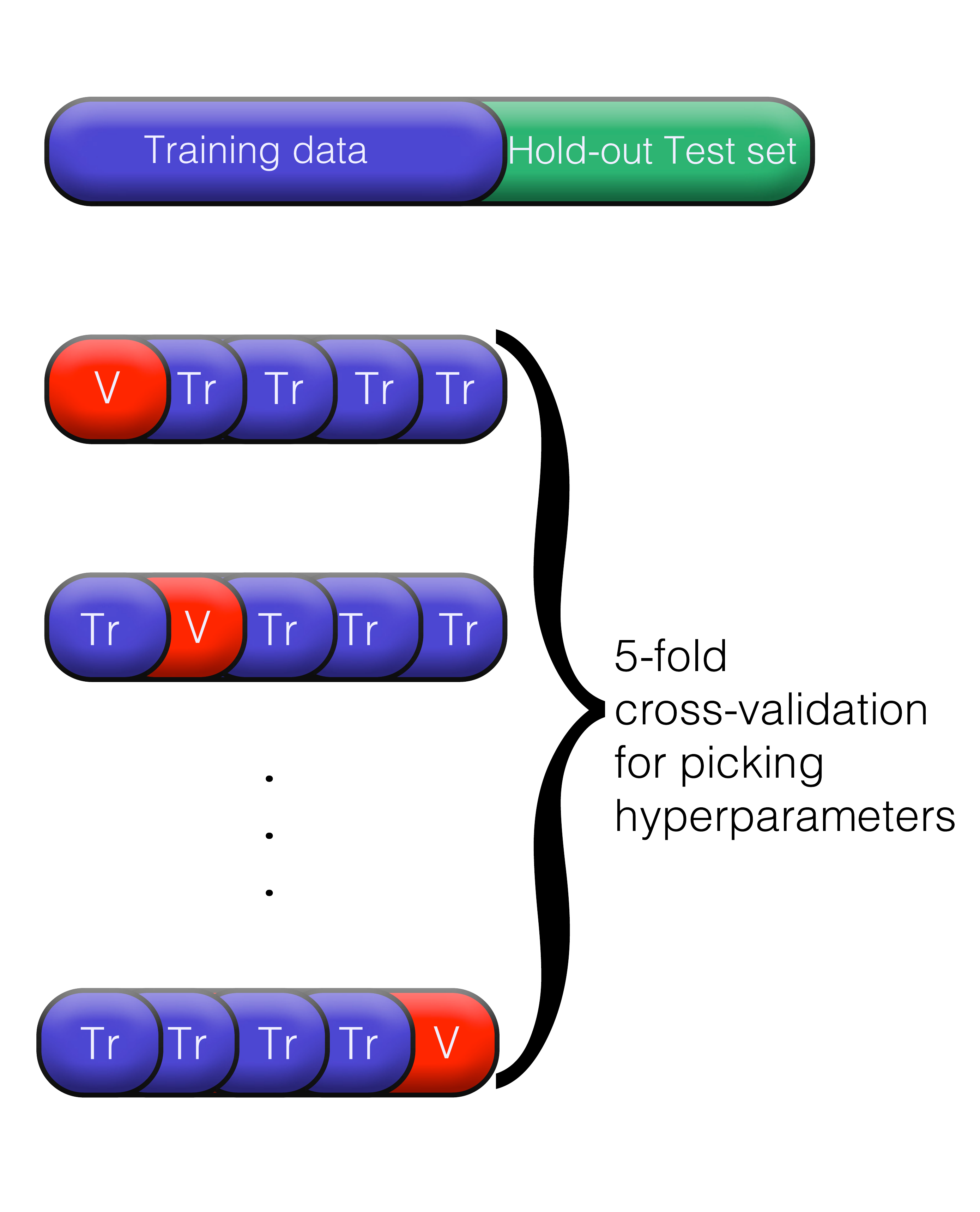}
  \vspace{-5mm}
  \caption{Representation of the dataset division: the training set is used for training, with cross-validation to select the various hyperparameters presented in \autoref{table:parameters}; the testing set is held out completely, and only used for the final model evaluation. \label{fig:training_diagram}}
\end{figure}

We designate $\Nlos=10^4$ lines to be a training set, and separate $10^4$ lines to be a testing set that will only be used for the final model evaluation. Then $\Npoints = \Nlos/N_g$ is the number of independent groupings used for training. The training set is further partitioned to perform 5-fold cross-validation (see \autoref{fig:training_diagram}) for selecting hyperparameters summarized in \autoref{table:parameters}. The total number of model parameters is listed in \autoref{table:total_parameters}.

Note that while the density transformation $t(\delta)$, grouping size $N_g$, and rescaling factor $c$ are choices that we make to optimize a particular model $\mathcal{M}:\mathcal{F}\to \tau$, the maximum wavenumber $\kmax$ directly influences what the model predicts. In an ideal situation, we would like to be able to produce a model that predicts values accurately for a large $\kmax$ because this would allow studies of a wide range of modes. However, as $k$ increases, non-linear evolution of the density field may break the connection between total matter density and properties of neutral hydrogen, and therefore, we expect that the model will not be able to produce accurate results. This is confirmed in our numerical experiments, where we observe that the model performance degrades as we increase $\kmax$.

While we employ a standard \textit{mean squared error (MSE)} as the training loss, we will use a customized error metric for cross-validation and testing, described in the next subsection. The training loss minimization is performed using the ADAM algorithm~\citep{Kingma2015}, with learning rate 0.001 and 1st and 2nd moment exponential decay rates of $(\beta_1,\beta_2) = (0.9,0.999)$ coupled with back-propagation and a batch size of 128 datapoints for each iteration. An early stopping criteria with patience of 15 epochs is used, and we recover the best weights attained during the training phase as the final model. Lastly, we also vary the fraction of training data used to explore the sensitivity with respect to training data size, and the amount of training data needed to attain a certain accuracy for our chosen DNN architecture.

\begin{table}
    \centering
    \caption{Scientific model hyperparameters. In addition to \textit{standard} DNN hyperparameters (e.g., network architecture, activation function, etc.) we are interested in finding the optimal configuration of our model with respect to how we represent the data.
    \label{table:parameters}}
    \begin{tabularx}{\columnwidth}{l|l}
 \hline
 Hyperparameter & Choices \\ \hline
 Density transformation $t(\delta)$ & $\delta$, $\ln(1+\delta)$, $1-(1+\delta)^{-1}$ \\[1mm]
 Grouping size $N_g$ & 1, 2, 5, 10, 20, 50, 100 \\[1mm]
 Rescaling factor $c$ in $P_{\rm norm}$ & $\sfrac{1}{4}$,$\sfrac{1}{2}$, $1$, $2$ \\[1mm]
 Max output $\kmax (\hMpc)$ & 1, 2, 4, 8 \\[1mm]  
 \hline 
    \end{tabularx}
\end{table}

\begin{table}
    \centering
    \caption{Number of free parameters (weights and biases) for the chosen model \autoref{eq:ann_model}. These scale with the input $\Nin$ and output size $\Nout$.
    \label{table:total_parameters}}
    \begin{tabularx}{\columnwidth}{l|l}
 \hline
 $\Nin, \Nout (\kmax)$ & Number of free parameters \\ \hline
 256, 31 (1) & 107,999 \\ \hline
 512, 63 (2)  & 432,063 \\ \hline
 1024, 127 (4) & 1,728,383 \\ \hline
 1024, 254 (8) & 1,931,905 \\[1mm]  
 \hline 
    \end{tabularx}
\end{table}

\subsection{Evaluating performance}
\label{sec:performance}

We evaluate the predictive performance of the model for the matter density power spectrum with the following error function:
\begin{equation}
 \label{eq:average_error}
 \mathcal{E}_0 = \frac{1}{N_{\rm out}} \sum_{k}^{\kmax} \frac{|\Pkpred-\Pktrue|}{\Pkpred+\Pktrue}
\end{equation}
where
\begin{equation}
 \Pk \equiv 
 \frac{1}{\Npoints}\sum_{l=1}^{\Npoints} t_{\mathrm{out},l}(k)
\end{equation}
is the spectrum coefficient corresponding to wavenumber $k$ averaged over all groupings in the test (or validation) set. $\Pktrue$ is the signal calculated directly from the simulation. This function is the symmetric mean absolute percentage error (SMAPE) between $\Pkpred$ and $\Pktrue$, which measures the relative errors between these quantities. Note that the number of terms in the sum will vary depending on $\kmax$.

We use this relative difference of the predicted and true power spectrum because it is the most basic measure of model performance. Perturbation modes in the quasi-linear regime, when $\Delta_1(k) \lesssim 1$, are independent of each other. If the model is unable to recover the average power spectrum in the quasi-linear regime, it would not be able to predict the power at different $k$ for individual lines or groups of lines. Taking the relative difference, normalized by $\Pkpred+\Pktrue$, allows us to minimize the influence of a single mode with the largest value of $\Pk$. 

We note that we employ an evaluation loss (for validation/testing) in \autoref{eq:average_error} that differs from the training loss. 

We select this evaluation loss to illustrate several ideas. 
First, one may choose the loss functions to reflect specific desirable properties of the model prediction. For example in this case, \autoref{eq:average_error} indicates the average error (over all validation points), but one may also target the variance, other moments, pointwise matching, etc. 
Second, penalizing the average is a more lenient metric compared to the MSE loss, and we wish to begin by evaluating the models with this easier-to-achieve statistic.
At the same time, retaining the MSE for training can help ensure the models to produce physically realistic outputs, and prevent fortuitous cancellations of wildly unrealistic predictions that may still yield good averages.
In the future, we will explore evaluation metrics for individual or small groups of LoS. Overall, both the training and evaluation losses may be customized to better align with the desired properties of the models. 
Typically, the training loss for a gradient based optimisation must be differentiable, and the performance metric should be easily interpretable.

\section{Results}
\label{sec:results}

We begin by addressing the choice of model hyperparameters from \autoref{table:parameters}. We select these hyperparameters by performing 5-fold cross-validation as illustrated in \autoref{fig:training_diagram} -- that is, finding the best hyperparameter setting offering the lowest average (over all folds) cross-validation loss based on \autoref{eq:average_error}. Approaching this multi-variate mixed-integer optimisation problem directly is very expensive; instead, we investigate the effects of one hyperparameter at a time while fixing all others, and progressively zoom in to their well-performing values. Our final findings, as well as results illustrating these choices, are summarised on \autoref{table:summary}. 

\begin{table}
    \centering
    \caption{Optimal choices of hyperparameters based on numerical experiments. \label{table:summary}}
    \begin{tabularx}
    {\columnwidth}{l|l|l}
 \hline
 Hyperparameter & Best Choices & Evidence \\ \hline
 Rescaling factor $c$ in $\Pnorm$ & 1 or $\sfrac{1}{2}$ &  \autoref{fig:res_select} \\[1mm]
 Density transformation $t(\delta)$ & $\ln(1+\delta)$ & \autoref{fig:density_transformation} \\[1mm]
 Grouping size $N_g$ & 5 or 10 & \autoref{fig:res_select}, \autoref{fig:density_transformation} \\[1mm]
  Max output $\kmax (\hMpc)$ & 1 or 2 &  \autoref{fig:res_select}, \autoref{fig:density_transformation} \\[1mm]  
 \hline 
    \end{tabularx}
\end{table}

\begin{figure}
    \includegraphics[width=\columnwidth]{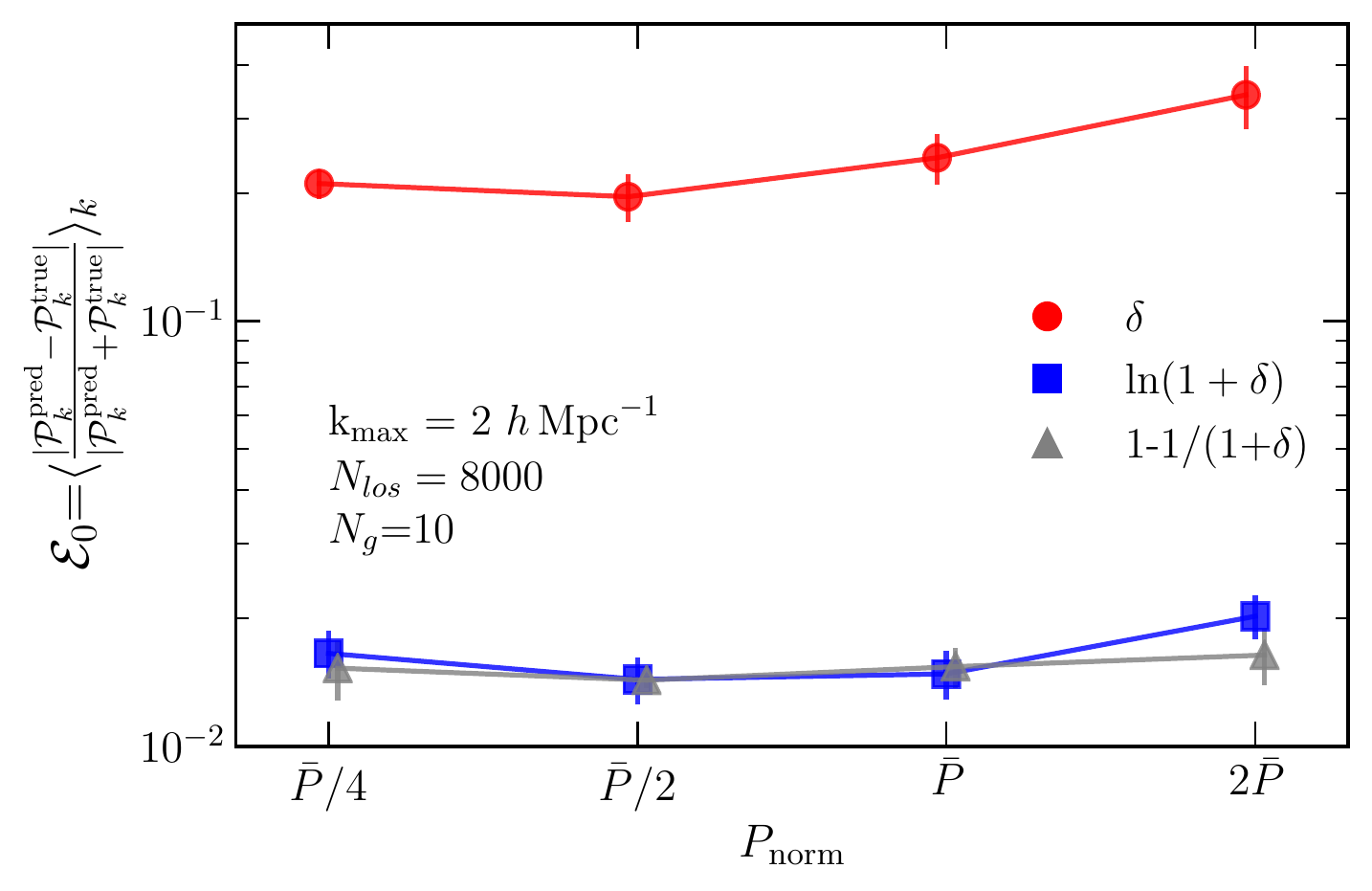}
    \vspace{-5mm}
    \caption{Selection of normalisation constant $\Pnorm$ for the model output, defined by \autoref{eq:normalisation_output}, with different choices of the output transformation $t(\delta)$. Using the original overdensity $\delta$ results in a much reduced accuracy of the model inference, as discussed in detail in Appendix. The logarithmic and inverse transformation give similar results. The results are robust to the choice of values of $\Pnorm$ in the considered range.}
    \label{fig:res_select}
\end{figure}

\autoref{fig:res_select} shows how the performance of the model changes with the normalisation factor $\Pnorm$ and density transformation. The results are computed from validation sets as the average over the 5 folds, with the error bar denoting plus/minus one standard deviation. From this plot, we first note that the logarithmic and inverse transformations perform very similarly, while the original overdensity $\delta$ (``linear transformation'') performs significantly worse. The normalisation factor $\Pnorm$ does not appear to influence the performance much for the logarithmic and inverse transformation models, but it has a greater effect for the linear transformation. This conclusion holds within the considered range of $\Pnorm$, and the error deteriorates noticeably as we move away from this range. Selecting for the lowest error, we thus set the normalisation to be $\Pnorm = \bar{P}/2$ (i.e. $c=\sfrac{1}{2}$) for the remaining numerical experiments in this paper. 

\autoref{table:ng_transformation} lists the values of $\mathcal{E}_0$ for the three choices of density transformation, and for grouping size $N_g$ from 1 to 100. The linear transformation performs the worst across all $N_g$, while the logarithmic and inverse transformations behave similarly; these trends are consistent with \autoref{fig:res_select}. The behavior can be further understood from \autoref{sec:appendix}, where we show that the linear transformation is most dominated by high density peaks, and therefore, expect the model to perform worse. At last, we choose the logarithmic transformation since it is more intuitive than the inverse transformation. 

\begin{figure}
  \includegraphics[width=\columnwidth]{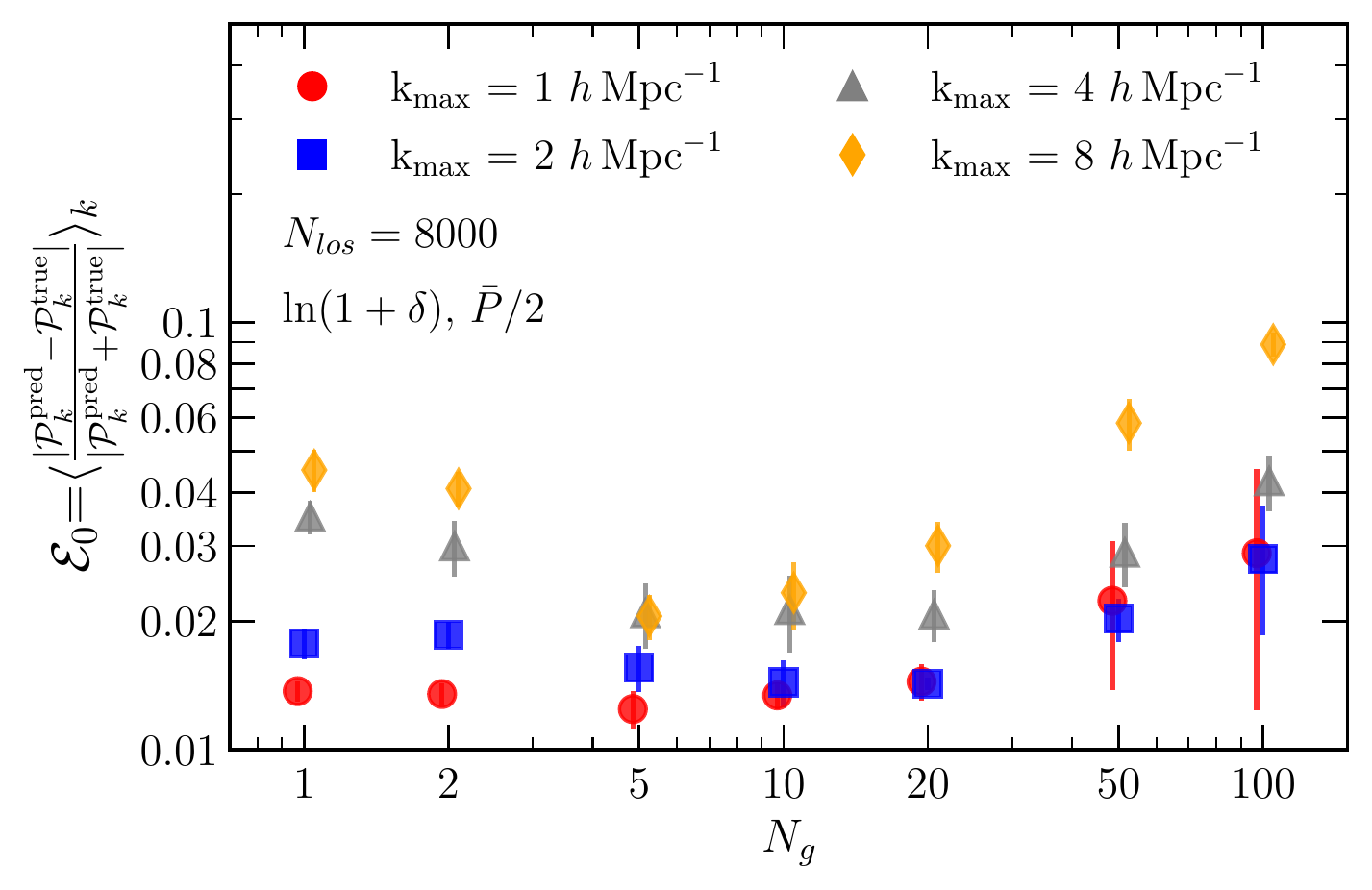}
  \vspace{-6mm}
  \caption{Model performance for different grouping size $N_g$ and maximum wavenumber $\kmax$. We show the best-case results with the largest data size $\Nlos = 8000$ and optimized values of the density transformation $t(\delta)$ and normalisation $\Pnorm$. Limiting the model to larger scales of $\kmax \le 2\hMpc$ results in better model performance, while the dependence on the grouping size is weak as long as $N_g < 50$.}
  \label{fig:density_transformation}
\end{figure}

In \autoref{fig:density_transformation}, we compare different grouping size $N_g$ and maximum wavenumber $\kmax$, while fixing output transformation and normalisation factor to be our best choices: $\ln(1+\delta)$ and $\Pnorm=\bar{P}/2$, respectively. We attain the best performance with $N_g = 5$ or $10$, although the results remain similar for any grouping size smaller than 50. The figure also shows how far into the small scales (corresponding to larger $k$) the density field can be probed: $\kmax = 1, 2\hMpc$ attains relatively low error, while $\kmax = 4, 8\hMpc$ models perform visibly worse. We thus take our default $\kmax$ to be $2\hMpc$ in this study, although the choice of $\kmax$ can also be influenced by the investigation context (e.g., if one is interested in probing a specific range of scales); see \autoref{sec:discussion} for more discussion.

\begin{table}
    \centering{
    \caption{Model performance factor $\mathcal{E}_0$ for different grouping size $N_g$ and output transformation $t(\delta)$, with fixed $\kmax=2\hMpc$ and $\Nlos = 8000$. \label{table:ng_transformation}}
    \begin{tabular}{c|c|c|c}
 \hline
 $N_g$ & $\delta$ & $\log(1+\delta)$ & $1-1/(1+\delta)$  \\
 \hline
1   & 0.118 & 0.018 & 0.020 \\
2   & 0.122 & 0.019 & 0.016 \\
5   & 0.169 & 0.016 & 0.013 \\
10  & 0.196 & 0.014 & 0.014 \\
20  & 0.188 & 0.014 & 0.015 \\
50  & 0.056 & 0.020 & 0.022 \\
100 & 0.065 & 0.028 & 0.024 \\
\hline
    \end{tabular}}
\end{table}

\begin{figure}
    \centering
    \includegraphics[width=\columnwidth]{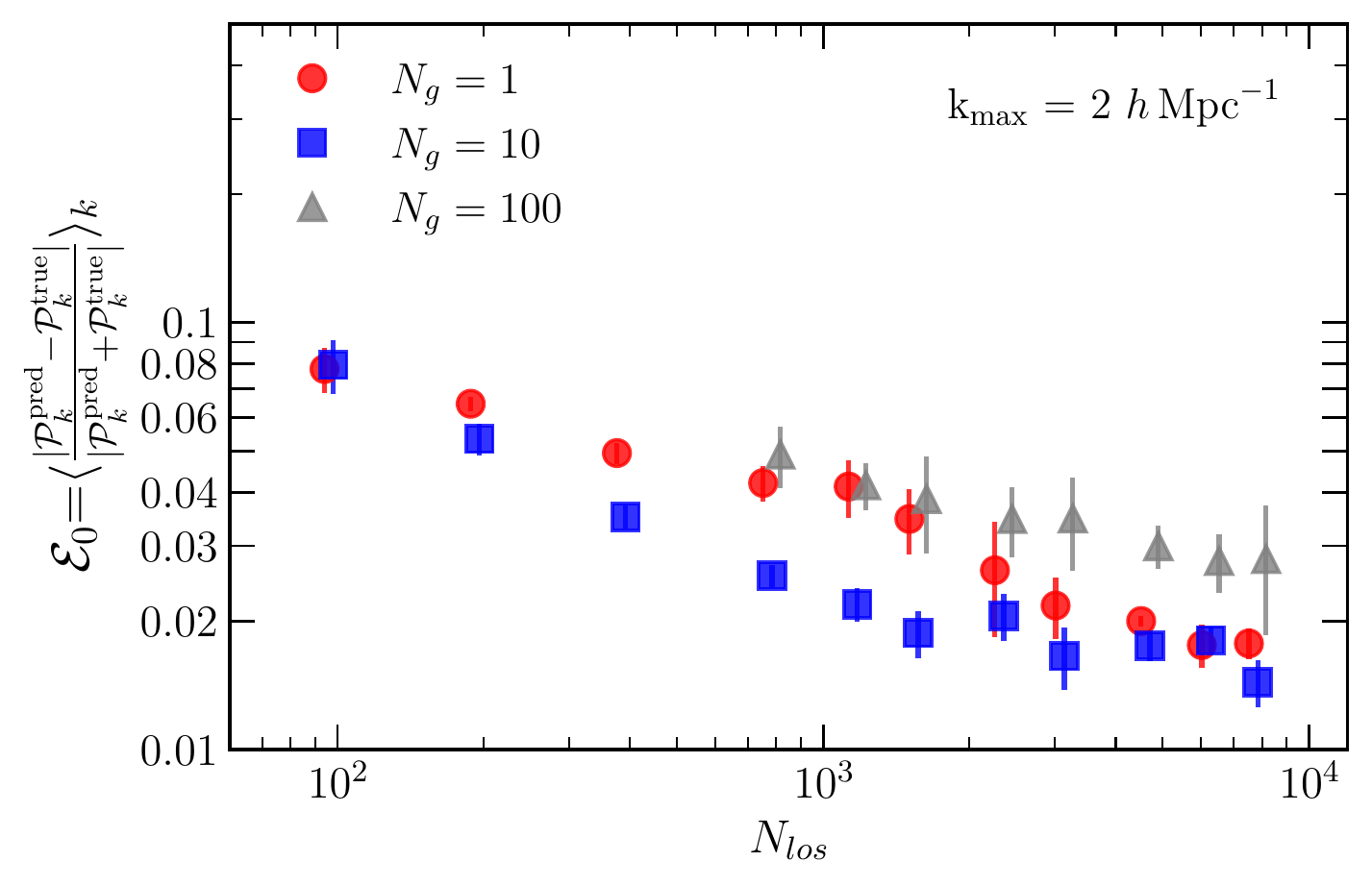}
    \vspace{-6mm}
    \caption{Accuracy of reconstruction of the logarithmic density transformation as a function of training set size $\Nlos$, for several choices of grouping size $N_g$. We observe that increasing the size of training dataset leads to an improved performance when the dataset sizes are small, however, as the dataset size becomes larger, the performance gains are reduced. After approximately 1000 LoS, the model's performance increases only marginally.}
    \label{fig:eps_resc_log_k1}
\end{figure}

In \autoref{fig:eps_resc_log_k1}, we show the error on the reconstruction of the power spectrum as a function of number of lines-of-sight considered in the training process ($\Nlos$), for $\kmax=2\hMpc$ and three choices of the grouping size $N_g$. As expected, the error decreases with the number of lines-of-sight in the training set. However, we can observe that after using a training dataset of a few thousand, the performance of the models appears to converge to a constant value of the error.

This trend suggests that for this task and with our model architecture and training selections, we do not require more than a few thousand data points to obtain convergence of the model's performance. If a larger dataset is available from an actual observation, the model complexity could be increased to match it. For a smaller dataset ($\Nlos \lesssim 1000$), \autoref{fig:eps_resc_log_k1} can be used to read off the expected value of error $\mathcal{E}_0$.

\begin{figure}
     \includegraphics[width=\columnwidth]{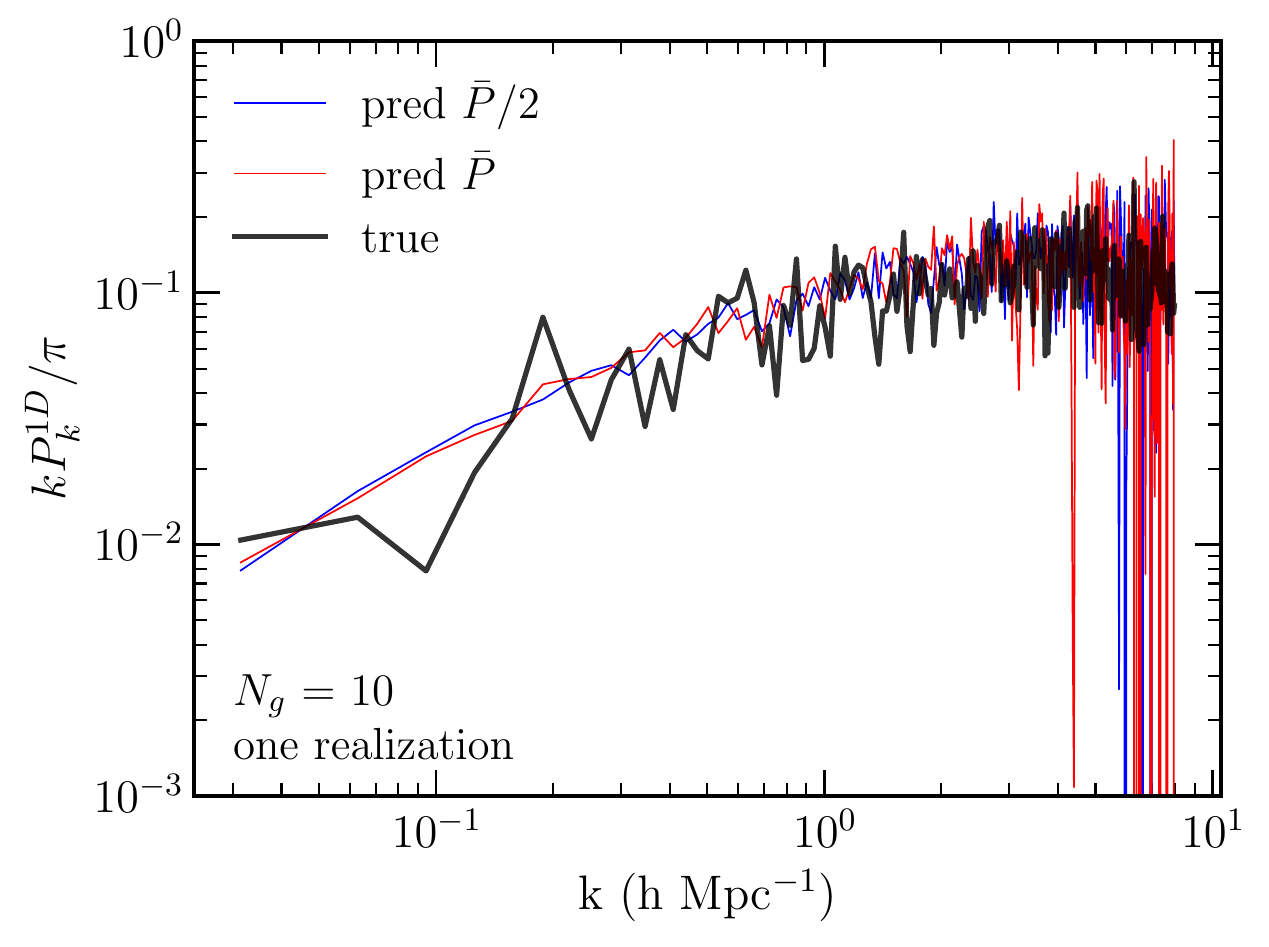}
     \includegraphics[width=\columnwidth]{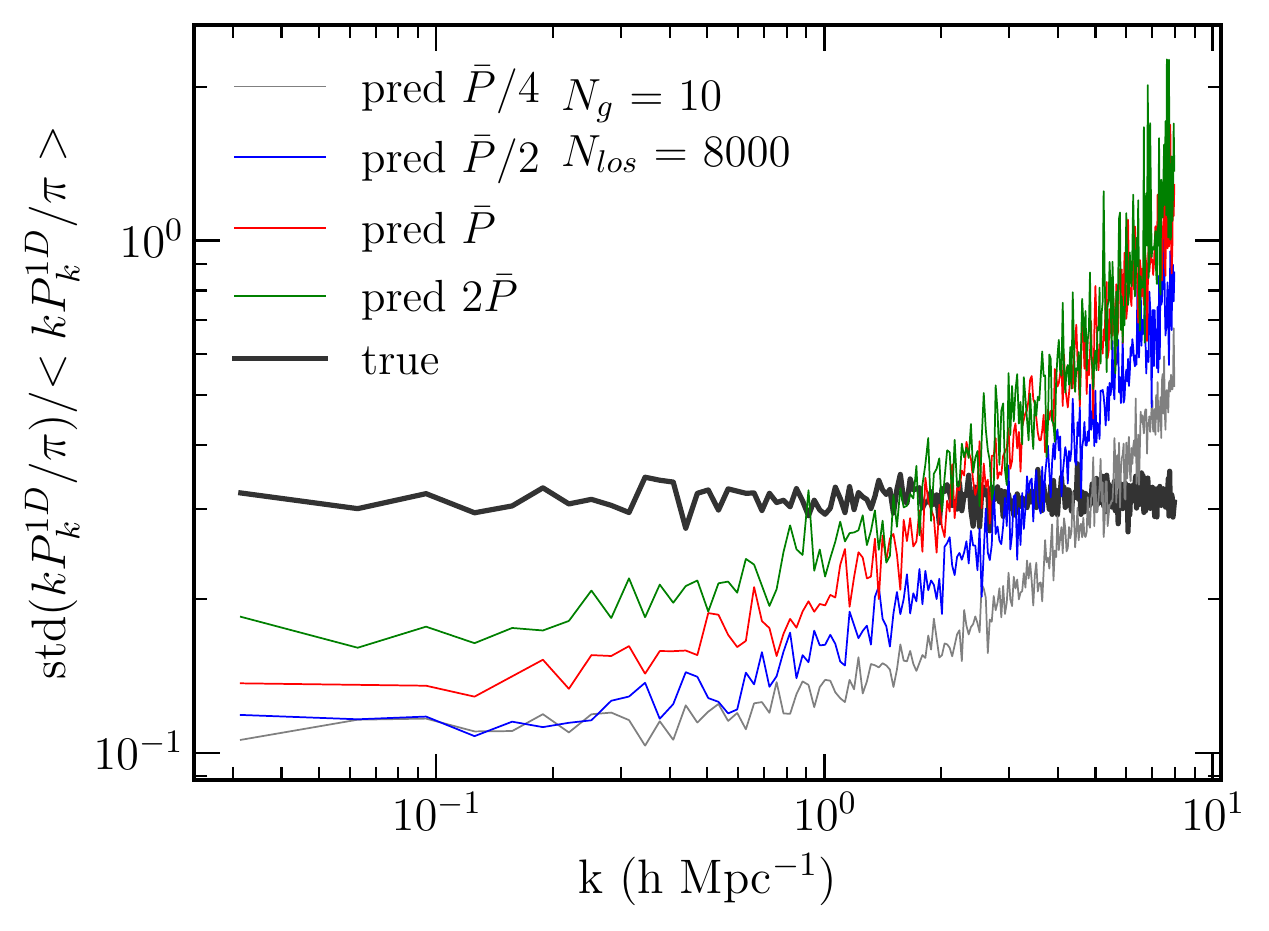}
    \vspace{-6mm}
    \caption{An illustrative example of the effect of rescaling the output in a model with $\Nlos = 8000$, $N_g = 10$, $\kmax = 8\hMpc$, and $t(\delta) = \ln(1+\delta)$.
    \textit{Top panel:} One validation point. We can observe that the model predicts less variance with $k$ on large scales and more variance on small scales, compared to the true simulation. Two models with different $\Pnorm$ produce similar, but not identical power spectra.
    \textit{Bottom panel:} Standard deviation of the predicted  power spectrum in the full validation dataset as a function of $k$. The amount of variance increases with the normalisation factor $\Pnorm$.}
    \label{fig:different_normalisations}
\end{figure}

From all these experiments, we conclude that the best choice of hyperparameters for this model is the following: $t(\delta) = \ln(1+\delta)$, $N_g = 10$, $c = \sfrac{1}{2}$ and $\kmax = 2\hMpc$. With this choice and using $\Nlos=3200$ to train our fiducial model, we take a new, completely unseen test set to evaluate the performance of our model. We find $\mathcal{E}_0 = 0.014$, very similar to the results shown in \autoref{table:ng_transformation}. This demonstrates that we obtain equal performance on unseen data.

Since our performance metric $\mathcal{E}_0$ evaluates models on the averaged value of power spectrum over many modes $k$, we can now study how well the fiducial model works as a function of $k$. In the top panel of \autoref{fig:different_normalisations} we show an example of reconstruction of the dimensionless mass fluctuation $k P_{k}^{1D}/\pi$ for one randomly selected validation point. On large scales ($k \lesssim 1\hMpc$) the model predicts smoother function of $k$, while on smaller scales it visibly predicts more variation with $k$. We address this issue further in \autoref{sec:discussion}.

The bottom panel of this figure shows that the amount of variance in predicted power shows that no choice of the normalization factor allows the model to match the amount of variance of the true simulated data at all $k$. We note that this comparison of the variance should not be interpreted as direct performance of the model, because the variance was not included in evaluation metric. Instead, it is a demonstration of additional results predicted by the model. In the future, if the pointwise variance is deemed an important quantity for us to capture, it may be explicitly incorporated into the training and evaluation loss functions.

\begin{figure}
    \centering
    \includegraphics[width=\columnwidth]{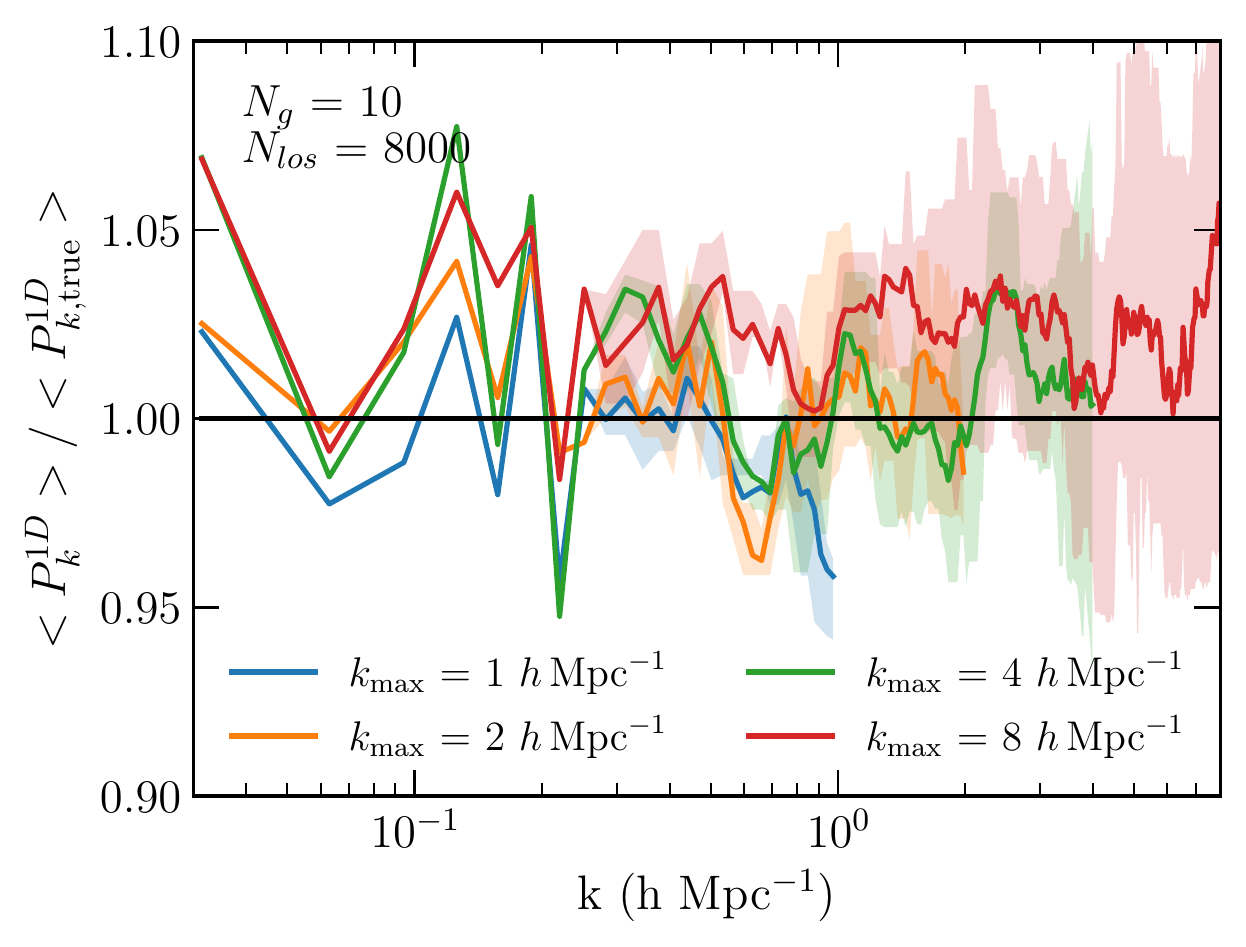}
    \vspace{-6mm}
    \caption{Deviation of the model output from true values of PS, for models with different choices of $\kmax$. On small scales $k > 0.2\hMpc$, the ratio is smoothed with a moving window in $\log{k}$ with the size of approximately 0.1~dex. For these scales solid lines show the median of the data points, while shaded regions show the 25\%-75\% interval of the moving window range. On larger scales the solid lines show the actual ratio.}
    \label{fig:ps_recon_diff_k}
\end{figure}

Finally, in \autoref{fig:ps_recon_diff_k}, we show direct comparison between the predicted and true power spectra, for several choices of $\kmax$. All other hyperparameters are kept fixed at their optimal choices: $t(\delta) = \ln(1+\delta)$, $c = \sfrac{1}{2}$, and $N_g = 10$. The ratio of the predicted to true power spectrum is close to unity at all scales and oscillates around a mean of $\sim 1$ as $k$ increases. On the logarithmic x-axis scale many modes are compressed at high $k$. To display the mean trend more clearly, we smooth the ratio with a moving window of 0.1~dex in $\log{k}$, for small-scale modes with $k> 0.2\hMpc$. The larger-scale modes are displayed without smoothing.

We can observe that as $\kmax$ increases, the error on the prediction also increases gradually. The median of all lines remains within 5\% of the true PS, but near $k\sim 8\hMpc$, individual lines deviate (both, over and under) by up to 10\% for the interquartile range and by a larger amount for other lines. This is consistent with the behavior of the performance metric $\mathcal{E}_0$ shown in \autoref{fig:density_transformation}, which shows more variance in the average predictions for higher values of $k$. Here we can verify that the source of error is dominated by the larger wavenumbers. This justifies our selection of only moderately non-linear scales, $\kmax=2\hMpc$, as our optimal model.

\section{Discussion}
\label{sec:discussion}
\subsection{Present Model}

Some of the parameters we consider, such as the grouping size $N_g$, density transformation $t(\delta)$, or scaling factor $c$ are genuine hyper-parameters, which can be fixed by optimizing the model performance. The two key parameters -- $\kmax$ and $\Nlos$ -- however, may instead be determined by the scientific problem in question and the size of observational sample. There are currently over 100 quasars known above $z=6$ \citep{banados2016,wang2019}, but the largest progress in increasing observational samples is expected after the next generation, 30-meter class telescopes become operational later in this decade. With the approximately 10 times increase in collecting area over the current largest existing telescopes, and the steep slope of the quasar luminosity function, $N(>L)\propto L^{-1.7}$, the next generation of quasar survey will increase observational samples by a factor of 30--50. This served as our motivation for considering $\Nlos$ in the range of a few thousand, even though we can generate much larger simulation datasets. Even that amount of observational data is probably a maximum limit, and therefore in future works, it would be useful to consider models that could be trained with less data.

The choice of $\kmax$ is largely problem-dependent. For example, for cosmological constraints $\kmax\sim 1\hMpc$ may suffice, as the matter power spectrum is currently calibrated to about 1\% precision to that scale and to 3\% precision to $\kmax\sim 5\hMpc$ \citep[][at $z=0$, but the accuracy at higher redshifts is expected to be comparable or even better due to weaker clustering on a given spatial scale]{ho2021,arico2021}. For other applications, pushing deeper into the non-linear regime may be desirable. For example, probing the gas power spectrum to $\kmax\sim 10\hMpc$ will be useful for comparing the clustering of the cosmic gas in the general intergalactic medium with that in quasar proximity zones \citep{chen2021}, as a way of constraining quasar lifetimes and observational biases.

For our particular problem in this paper, we had to make a number of choices in setting up and training the neural networks. Main constraints were the computing time required to train each neural network and a large physical hyperparameter space to explore, in addition to the hyperparameters associated with the neural network model (size, activation function, regularisation, number of epochs, loss function). Such a large space of hyperparameters also made it prohibitively expensive to use a standard hyperparameter optimizer such as Hyperopt \citep{hyperopt}.
Some of the lessons that we learned in the process of working on this problem were:
(i) the importance of scaling the values of input and output (\autoref{sec:network}); (ii) monitoring for model over-fitting and stopping the training early to minimize its effects; and (iii) keeping in mind that the number of gradient updates per epoch varies as we vary the dataset size, which may require adjustment to the early-stopping criteria.

One issue that became clear in our current results is the inability of the model to recover the full variance of the dataset (\autoref{fig:different_normalisations}). Relative to the true power spectrum, our best model predicts smoother function of $k$ on large scales ($k \lesssim 1\hMpc$), while on smaller scales it predicts visibly more variance with $k$. The former is expected -- in the linear regime (on large spatial scales) individual $k$ modes are independent and hence uncorrelated. Since in the current implementation we use all the modes as input, the neural net generically introduces correlations among them. The model does not automatically remove such spurious correlations because the training loss does not penalize for their presence. Alternatively, one can imagine training a separate model for each $k$ mode, but such a setup would not account for the real physical correlations between the modes in the mildly non-linear regime that are included in our current model. The latter is important and allows our best model to recover the power spectrum even for $k \gg 1 \hMpc$.

One may consider using data-driven models which can represent more complex mappings, for example, through convolutional neural networks (CNN), to fix the lack of variance on large scales. However, a convolution of neighboring data points introduces a correlation between them. In addition, the modes on small scales are indeed strongly correlated, but the mode correlation length is strongly wavenumber dependent (becoming zero in the linear regime of large scales), and such a strongly variable correlation between modes does not easily decompose into a series of convolutions. Hence, it appears that the only way to both explicitly eliminate correlations between $k$ modes in the linear regime and to allow for such correlations in the mildly non-linear regime is to include an appropriate penalty in the training loss. It is not clear yet how to achieve this without prior knowledge of the signal to be recovered.

One can also consider how observational effects such as readout noise or instrumental resolution, absent in our synthetic input spectra, may affect the power spectrum recovery. In this exploratory work we did not consider such observational effects; since they are all important on small scales, we do not expect them to affect sufficiently small wavenumbers. One effect that does affect large scales is the quasar continuum placement. Modern, PCA-based continuum placement models are highly accurate \citep{bosman2021}, but not so accurate as to make the quasar continuum errors negligible. The bias in the continuum placement is, fortunately, estimated to be below the precision of our power spectrum recovery procedure, and hence the continuum placement errors can be averaged out to below that precision with a rather modest number of LoS (a few hundred).

\subsection{Future work}

In this work we considered the task of predicting the average density power spectrum from normalized quasar absorption spectra. As a next step, we are interested in predicting individual LoS quantities from the observed spectra. In this case, the model takes in a representation of one-dimensional quasar spectra data (for example, through Fourier modes), and predicts a one-dimensional representation of the density $\rho$. These pairwise quantities can be considered a \emph{time-series} (with the variable of ``time'' in this case being distance along the line), using one to predict the other. In this problem, other network architectures, such as recurrent neural networks (RNNs), might be more suitable to encode the variance associated with individual LoS.

There are additional issues to consider for this problem. For example, the quasar light is fully absorbed in regions that are dense enough, producing saturated absorption, making no model able to accurately predict the density in those regions. Thus, quasar spectra only contain information about regions with sufficiently low densities. This limitation can be mitigated if our ability to produce a trustworthy prediction is expressed through model uncertainty, for example, using Bayesian neural networks \citep{bnn_uncertainty}.

With the degeneracy of high density regions in mind, we can consider a classification problem instead, where we classify regions along the LoS simply as high, medium or low density regions. This is equivalent to a binned representation of the density field.

Other promising directions include probing even smaller scales ($k > 10\hMpc$) to investigate quasar proximity zones, as well as accounting for the effect of ionizing radiation escaping from massive galaxies near a given LoS.

\section{Conclusions}
\label{sec:conclusion}

We generate a novel dataset of pairs of 1-dimensional density fields and the correspondent absorption spectra from the CROC suite of galaxy formation simulations. We build an end-to-end methodology to infer the matter density power spectrum from the quasar spectra. We explore various ways to describe the data and set-up the model in order to optimize the metric specified in \autoref{eq:average_error}. With the best set of hyperparameters, we are able to reconstruct the power spectrum to an accuracy near 1\% up to $\kmax = 2\hMpc$. As $\kmax$ increases, the scatter produced by the model increases.

In future work, we would explore more challenging tasks, such as predicting the 1-dimensional density field from observed spectra and including full observed quasar spectrum. This work provides a foundation for developing advanced analysis methods for upcoming observations with JWST and 30-meter class ground-based facilities.

\section*{Acknowledgments}
We thank the Michigan Institute for Data Science for support in the form of the Propelling Original Data Science (PODS) grant. OG and XM were supported in part by the U.S. National Science Foundation through grant 1909063. MHV was supported by MIDAS. This manuscript has been co-authored by Fermi Research Alliance, LLC under Contract No. DE-AC02-07CH11359 with the U.S. Department of Energy, Office of Science, Office of High Energy Physics. This work used resources of the Argonne Leadership Computing Facility, which is a DOE Office of Science User Facility supported under Contract DE-AC02-06CH11357. An award of computer time was provided by the Innovative and Novel Computational Impact on Theory and Experiment (INCITE) program. This research was also part of the Blue Waters sustained-petascale computing project, which was supported by the National Science Foundation (awards OCI-0725070 and ACI-1238993) and the state of Illinois. Blue Waters was a joint effort of the University of Illinois at Urbana-Champaign and its National Center for Supercomputing Applications. 

\section*{Data availability}
The data underlying this article will be shared on reasonable request to the corresponding author.

\bibliographystyle{mnras}
\bibliography{references}

\appendix

\section{Data transformation}
\label{sec:appendix}

\begin{figure}
  \centering
  \includegraphics[width=0.94\columnwidth]{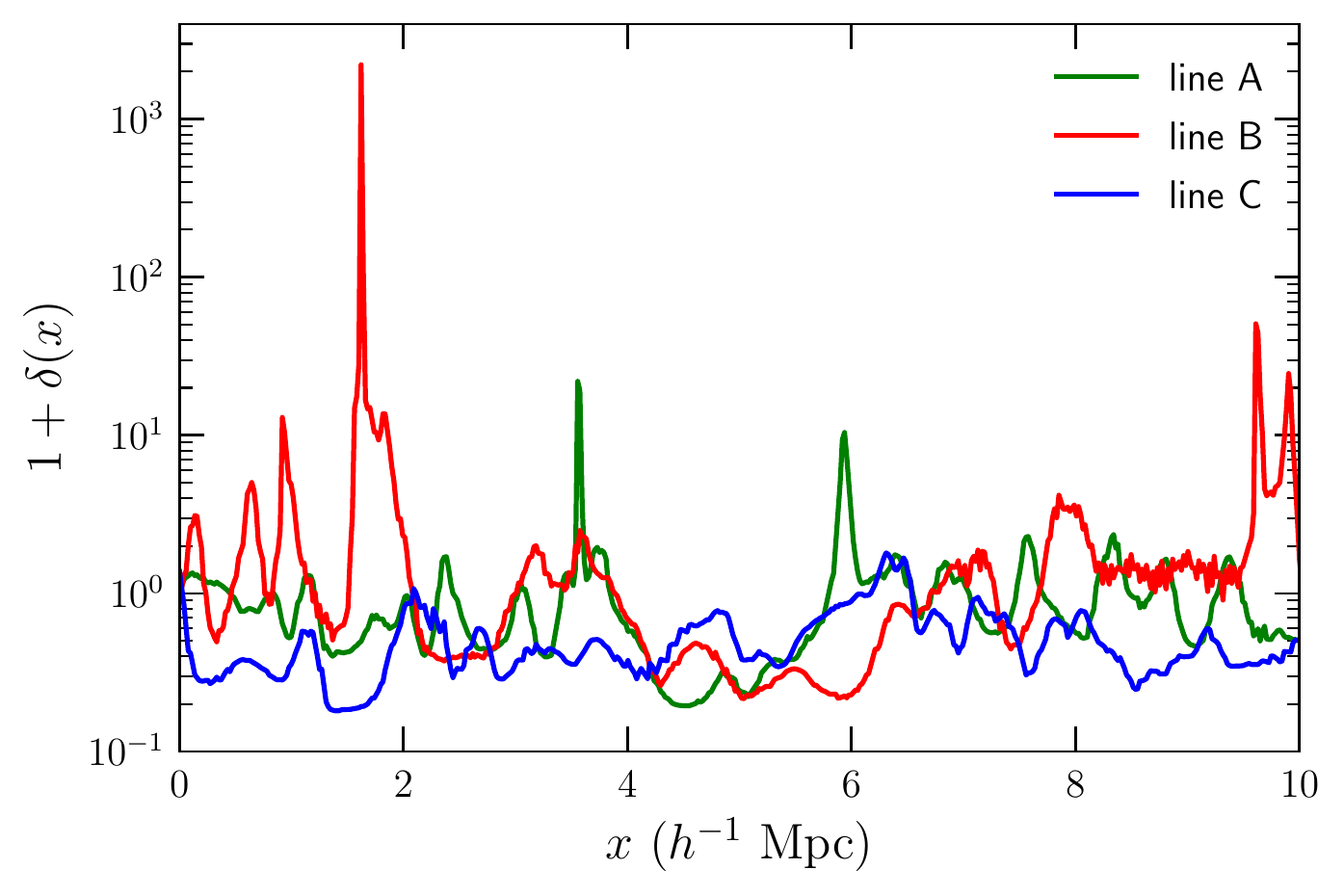}
  \includegraphics[width=0.94\columnwidth]{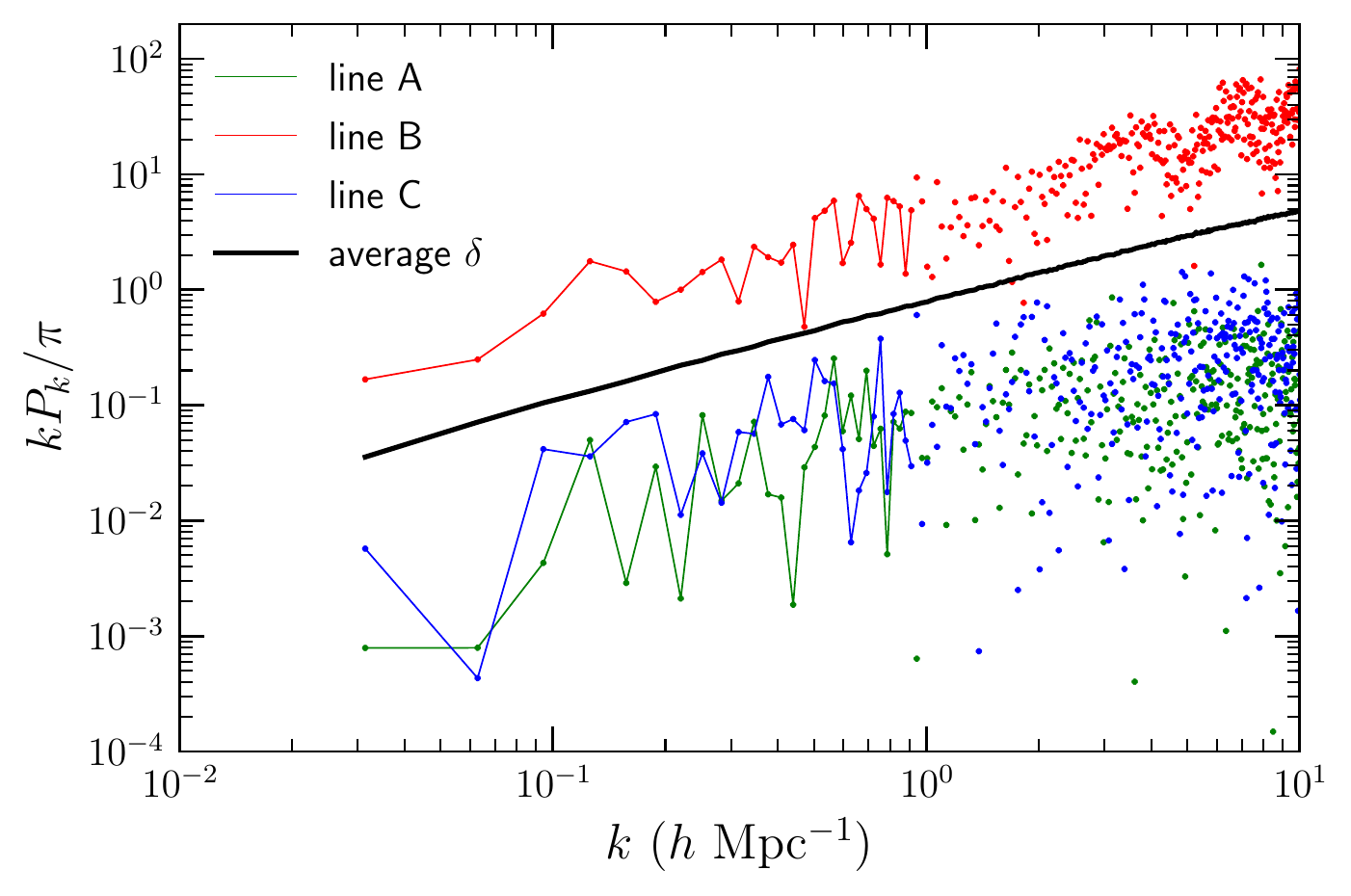}
  \includegraphics[width=0.94\columnwidth]{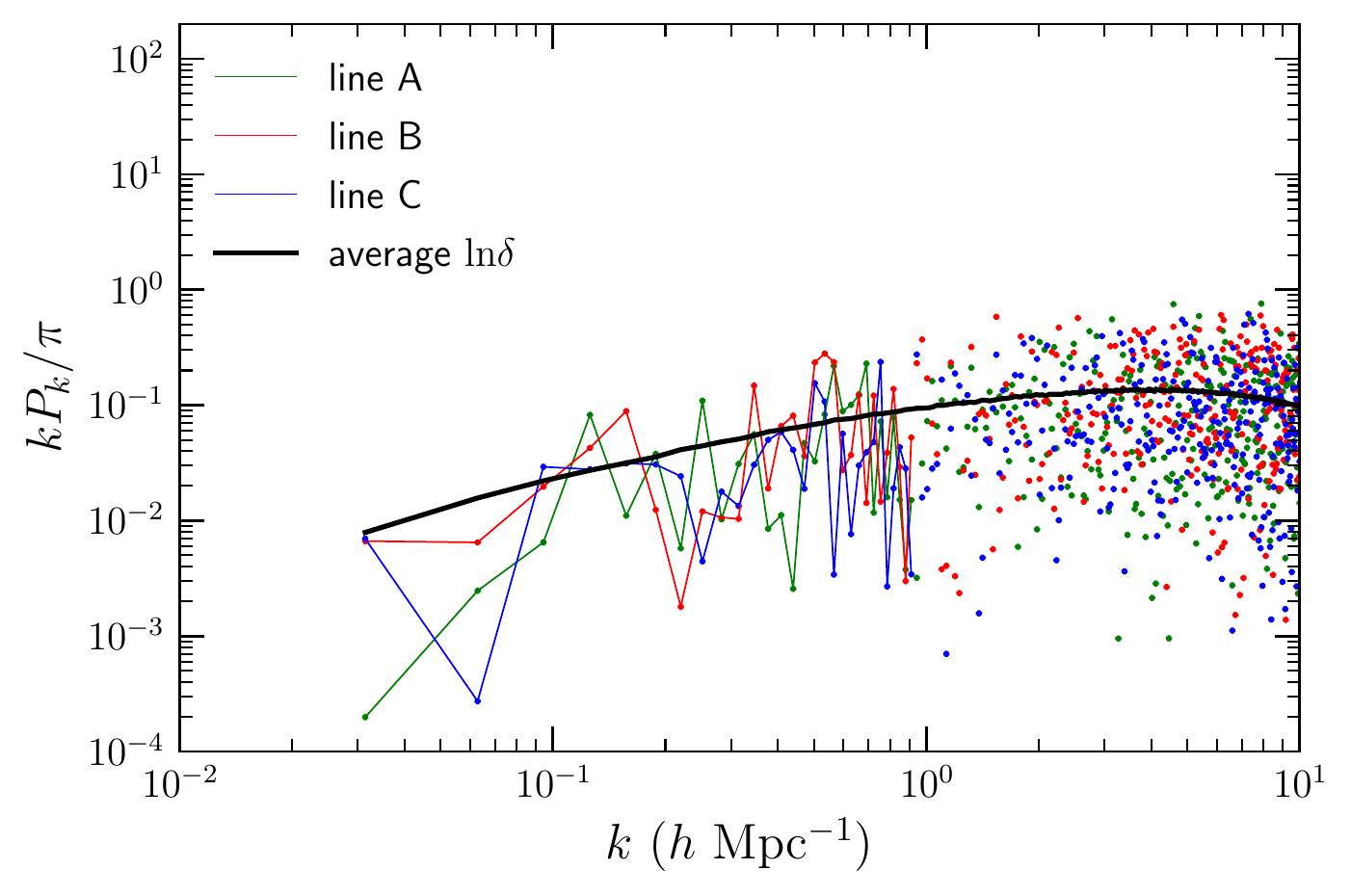}
  \includegraphics[width=0.94\columnwidth]{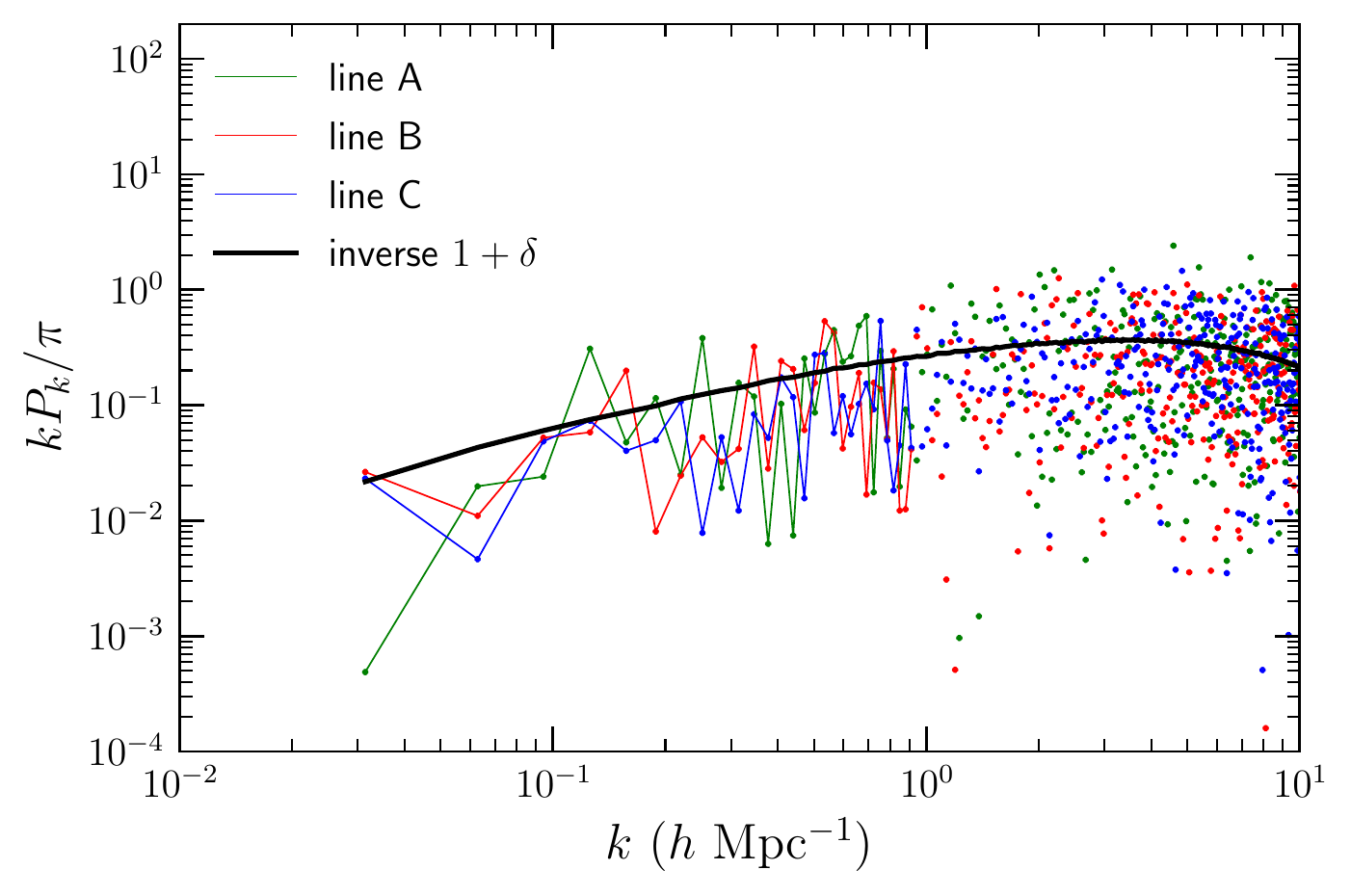}
  \vspace{-2mm}
  \caption{\textit{Top panel:} Illustration of the matter overdensity along a short part of 3 randomly chosen lines of sights. Line B has several large spikes, which bias the 1D PS. \textit{Bottom three panels:} The power spectrum of $\delta$, $\ln{\delta}$, and inverse $1+\delta$ for each of the three lines. Because of the large scatter of points, for clarity we do not connect them at high $k$. Black lines show the corresponding average of all lines of sight in the simulation.}
  \label{fig:fourier_comparison}
\end{figure}

The PS for individual LoS may differ significantly because of a few high-density regions dominating the total. The transformations of density field described in \autoref{ss:data_transform} are designed to mitigate their effects.

\autoref{fig:fourier_comparison} illustrates how the logarithmic and inverse transformations can significantly reduce the variation and range of individual lines. We use an example of three lines chosen randomly from the training dataset. The top panel shows the matter overdensity along a small part (5\%) of the line length, for clarity. Line B is significantly different from lines A and C in that it displays a very large (and rare) peak of overdensity $\delta > 10^3$. This small-scale peak leads to a much higher normalization of the original 1D PS at all wavenumbers (see second panel), relative to the average over all 100,000 lines, because of the integration over small scales (see \autoref{eq:pk1d}). On the other hand, the PS of lines A and C happen to be systematically lower than the average. Thus the PS of any of these lines are not representative of the cosmic average.

In contrast, the logarithmic transformation brings the resulting PS of all three lines close to each other and to the cosmic average (third panel of \autoref{fig:fourier_comparison}). The inverse transformation behaves similarly (bottom panel of \autoref{fig:fourier_comparison}). Therefore, either of these transformations allows us to use even a relatively small number of LoS to obtain a representative measure of the matter PS. For most of our results, we choose the logarithmic transformation, as it has a more intuitive interpretation.

\bsp
\label{lastpage}
\end{document}